\documentclass[10pt, conference, compsocconf]{IEEEtran}
\IEEEoverridecommandlockouts
\usepackage{cite}
\usepackage{amsmath,amssymb,amsfonts}
\usepackage{algorithmic}
\usepackage{graphicx}
\usepackage{textcomp}
\usepackage{xcolor}
\usepackage{xspace}
\usepackage{listings}
\usepackage{url}
\usepackage{balance}
\usepackage{booktabs}

\newcommand{\changed}[1]{{\color{black}#1}}

\newboolean{showcomments}
\setboolean{showcomments}{true}         
\ifthenelse{\boolean{showcomments}}
  {\newcommand{\nb}[2]{
  \fbox{\bfseries\sffamily\scriptsize#1}
     {\sf\small$\blacktriangleright$\textit{\textcolor{red}{#2}}$\blacktriangleleft$}
   }
  }
  {\newcommand{\nb}[2]{}
   
  }

\newcommand\tool{\textsc{VATE}\xspace} 
\newtheorem{definition}{Definition}

\begin{document}

\title{A Framework for In-Vivo Testing of Mobile Applications}

\author{
	\IEEEauthorblockN{Mariano Ceccato\IEEEauthorrefmark{1},
		Davide Corradini \IEEEauthorrefmark{2},
		Luca Gazzola\IEEEauthorrefmark{3},
		Fitsum Meshesha Kifetew\IEEEauthorrefmark{2}\\
		Leonardo Mariani\IEEEauthorrefmark{3},
		Matteo Orr\`u\IEEEauthorrefmark{3} and
		Paolo Tonella\IEEEauthorrefmark{4}}\\

  \IEEEauthorblockA{\IEEEauthorrefmark{1}\textit{University of Verona}, 
    Verona, Italy \\
  \IEEEauthorrefmark{2}\textit{Fondazione Bruno Kessler (FBK)}, 
    Trento, Italy \\
  \IEEEauthorrefmark{3}\textit{Universit\`a di Milano-Bicocca}, 
    Milan, Italy \\
  \IEEEauthorrefmark{4}\textit{Universit\`a della Svizzera Italiana}
    Lugano, Switzerland \\
  mariano.ceccato@univr.it,
  corradini@fbk.eu, 
  luca.gazzola@unimib.it,
  kifetew@fbk.eu,\\
  leonardo.mariani@unimib.it, matteo.orru@unimib.it,
  paolo.tonella@usi.ch
	}
}

\maketitle

\begin{abstract}
The ecosystem in which mobile applications run is highly heterogeneous and configurable. All layers upon which mobile apps are built offer wide possibilities of variations, from the device and the hardware, to the operating system and middleware, up to the user preferences and settings. Testing all possible configurations exhaustively, before releasing the app, is unaffordable. As a consequence, the app may exhibit different, including faulty, behaviours when executed in the field, under specific configurations. 

In this paper, we describe a framework that can be instantiated to support in-vivo testing of a mobile app. The framework monitors the configuration in the field and triggers in-vivo testing when an untested configuration is recognized. Experimental results show that the overhead introduced by monitoring is unnoticeable to negligible (i.e., 0-6\%) depending on the device being used (high- vs. low-end). In-vivo test execution required on average 3s: if performed upon screen lock activation, it introduces just a slight delay before locking the device.

\end{abstract}



\section{Introduction} \label{sec:intro}

Mobile applications are built to operate on a plethora of devices, each running a different version of the operating system and offering different hardware resources (screen resolution, sensors, etc.). Moreover, user preferences can affect various aspects of such applications, from their visual appearance to the enabled/disabled functionalities. Software testing is expected to check the behaviour of mobile apps in all possible configurations. However, this is practically impossible because both the number of combinations is exponential and some configurations are difficult to reproduce during pre-release testing.

Previous works~\cite{LuPZ0L19,GazzolaMPP17} show that different configurations may lead to a different coverage of the code and that some faults are ``field-intrinsic'', that is, they are inherently difficult to detect in-house. \changed{In particular,} Lu et al.~\cite{LuPZ0L19} developed a technique to identify the test cases whose behaviour is affected by the user preferences. They found that some faults are exercised and exposed only under very specific preference configurations. Gazzola et al.~\cite{GazzolaMPP17} report an empirical investigation of field failures. They observed that the main reason for the leakage of faults from  pre-release  testing to field usage is \textit{combinatorial explosion}. i.e., the huge number of configurations in which the software should be tested to expose faults that otherwise might give raise to field failures.

Mobile apps have often a very large user base that exercises the software under various configurations. Moving part of the testing activity to the field is therefore an appealing option to deal with the combinatorial nature of configuration-specific mobile app faults. However, the overhead introduced by such form of testing, which we call ``in-vivo'' testing, should be minimized, to make it acceptable for the end user.

In this paper we propose a model to represent the configuration space of a mobile app and we describe \tool, a framework that we developed for the Android operating system, which supports in-vivo monitoring and testing of new app configurations. Our framework resorts to managed profiles~\cite{managedProfiles} to isolate the in-vivo testing session from the normal user session. We measured the impact of configuration monitoring on low-end to high-end devices and found that the runtime increase is between imperceptible to negligible (on average between 0\% and 6\% CPU load overhead). Assuming that the actual execution of in-vivo tests takes place when the device is not under active usage (e.g., when it transitions to the screen lock mode), each test case is expected to introduce a delay of about 3 seconds on average. 

\tool is the first attempt to bring some testing activities to the field for mobile apps. The preliminary results obtained on a benchmark of six Android apps show that \changed{\tool} is a promising approach and that its impact on the end-user can be acceptable. The \tool tool and the experimental material are publicly available online~\cite{tool}.

\begin{figure*}[tb]
	\centerline{\includegraphics[width=0.75\textwidth]{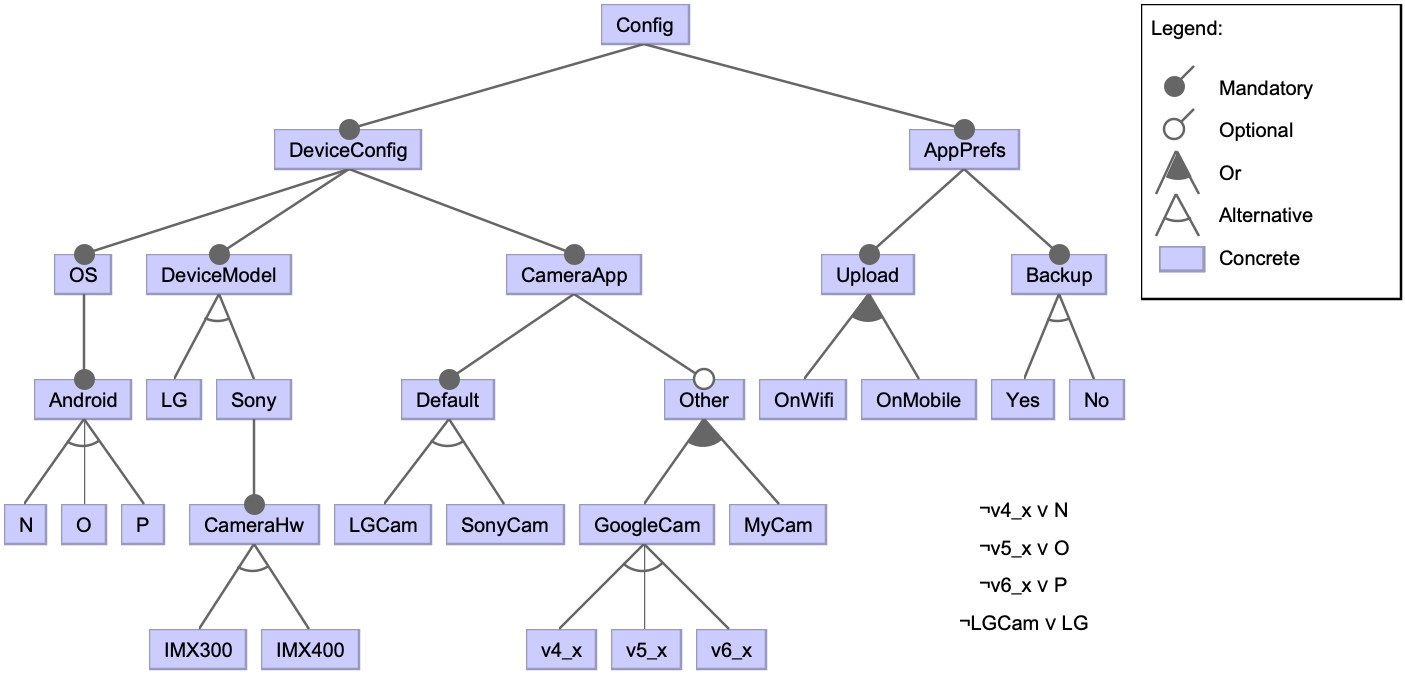}}
	\caption{Full configuration model for the \textit{ChatApp} application  } 
	\label{fig:chatapp_fm}
\end{figure*}

\section{Motivating Example} \label{sec:approach}


Let us consider a hypothetical messaging app for Android devices, which we call \textit{ChatApp} (pronounced shut-up). \textit{ChatApp} supports the exchange of messages and multimedia content between its users. Moreover, \textit{ChatApp} can take a picture of the user when the user creates or updates her\changed{/his} profile. To take a picture, \textit{ChatApp} sends an intent (see \changed{Listing~\ref{code:intent}}) to delegate the task to any app that can take a picture using the camera of the mobile device.

\begin{figure}[htb]
\begin{lstlisting}[language=Java, label={code:intent}, caption={Intent sent by \textit{ChatApp} to obtain a profile picture}, captionpos=b]
Intent cameraIntent = new Intent(MediaStore.ACTION_IMAGE_CAPTURE);
cameraIntent.putExtra(MediaStore.EXTRA_OUTPUT, outputImgUri);
startActivityForResult(cameraIntent, REQUEST_IMAGE_CAPTURE);
\end{lstlisting}
\end{figure}

Since \textit{ChatApp} relies on external resources (installed camera app; camera hardware) for the successful execution of the add/update profile image functionality, the scenarios in which a failure might occur depend on multiple factors: the hardware installed in the device, since the interaction with some camera models may fail; the configuration of the environment and operating system, since not all camera apps might be compatible with the ChatApp application; the settings of the app itself, since some specific choices might be not well supported by the app; and a combination of all these factors. Hence, adequately testing ChatApp requires addressing the combinatorial exploration resulting from all these factors.  

\textbf{\emph{Environment configuration}}. We use feature models~\cite{foda1990} to represent and manage the large configuration space that may affect apps. 
The configuration model of \textit{ChatApp} is shown in Figure~\ref{fig:chatapp_fm}, where inner nodes represent features; leaf nodes represent feature values; and the parent-child edges represent the feature-subfeature decomposition. While the default interpretation of feature decomposition is AND-decomposition, modifiers are available to express OR/XOR-decompositions and to identify a feature as mandatory/optional (see Legend in Figure~\ref{fig:chatapp_fm}). The logical constraints at the bottom-right are added to further constrain the admissible configurations.

The configuration  of \textit{ChatApp} is decomposed into two main parts: 1) \emph{DeviceConfig}, representing the configuration of the device on which the app is running; and 2) \emph{AppPrefs}, representing the various settings of the app itself.  \emph{DeviceConfig} includes the Android version (\textit{OS} feature), the camera apps that can be delegated \changed{to take pictures} (\textit{CameraApp}), and the actual model of the device (\textit{DeviceModel}), all of which are mandatory features. In turn, \textit{CameraApp} can be the default app (\textit{Default}, mandatory feature) or an additional app (\textit{Other}, optional feature). \textit{Default} can be instantiated by a set of mutually exclusive apps (empty arc), while \textit{Other} can be instantiated by a set of non exclusive apps (filled arc). When the device model is \changed{\textit{Sony}}, the camera hardware (\textit{CameraHw} feature) can be either \textit{IMX300} or \textit{IMX400}.

\emph{ChatApp}  has also a couple of application-specific settings. The first one (\emph{Upload}) represents a preference of the user to upload photos over wifi, mobile data, or  both. The other setting (\emph{Backup}) represents the preference of the user on whether or not to backup chats. The feature model contains also a few cross-tree constraints of type ``implies''. For instance, the cross-tree constraint (v4\_x $\Rightarrow$ N, equivalently shown as $\neg$4\_x $\lor$ N in Figure~\ref{fig:chatapp_fm}) indicates that version 4\_x of \textit{GoogleCamera} constrains the version of \textit{Android} to be N (\textit{Nougat}); the camera app \textit{SonyCamera} constrains the device model to be \textit{Sony}.


In addition to representing the full configuration space, we need to record also the set of configurations that have been tested so far. Let us consider \emph{ChatApp} at the time it is first deployed to its users and let us assume that pre-release testing has been carried out on an LG phone with default camera on all three Android versions, with user settings specifying that upload is possible only on the wifi and that backup is disabled. The set of tested configurations will include the following tuples of feature values:

\vspace{0.2cm}
\begin{footnotesize}
	\noindent
	$\langle$\textit{N}, \textit{LG},  \textit{LGCam}, \textit{OnWifi},  \textit{No}$\rangle$ \\
	$\langle$\textit{O}, \textit{LG},  \textit{LGCam}, \textit{OnWifi},  \textit{No}$\rangle$ \\
	$\langle$\textit{P}, \textit{LG},  \textit{LGCam}, \textit{OnWifi},  \textit{No}$\rangle$ 
\end{footnotesize}
\vspace{0.2cm}


\textbf{\emph{In-vivo testing of ChatApp}}. \tool includes a run-time in-vivo test component that can monitor the configuration elements relevant to the app and checks whether the current configuration is tested, untested or unknown.
%
%
This information can be extracted by a run-time probe that queries the device and the app preferences and compares the retrieved information to the tuples of tested configurations.

The following are examples of \textit{tested}, \textit{untested} and \textit{unknown} configurations of \textit{ChatApp}

\begin{footnotesize}
	\begin{description}
		\item [tested] ~~~$\langle$\textit{N, LG, LGCam, OnWifi, No}$\rangle$ 
		\item [untested] ~~~$\langle$\textit{N, Sony, SonyCamera, v4\_x, IMX400, OnWifi, Yes}$\rangle$ 
		\item [unknown] ~~~$\langle$\textit{P, Xiaomi, XiaomiCamera, Xiaomi/Dual camera, v6\_x, OnWifi, OnMobile, No}$\rangle$ 
	\end{description}
\end{footnotesize}

Different configurations trigger different reactions. A tested configuration triggers no reaction.  An untested configuration, triggers in-vivo test execution.  An unknown configuration triggers a feedback to testers who are asked to extend the model to incorporate the new cases that were not considered at the beginning, when the full configuration model was produced. In addition, an unknown configuration can be immediately validated with the available test cases.

Let us now consider the following hypothetical field failure:

\smallskip
\noindent	\textit{A new camera app, \textit{XiaomiCamera}, is installed. The camera hardware is deployed with a driver that, under Android version \textit{N}, does not initialize the camera if not requested explicitly. When \textit{ChatApp} takes a picture of the user, the request goes through \textit{XiaomiCamera}, which does not explicitly initialize the camera when responding to an intent (it initialises the camera only when activated by the user). Correspondingly, \textit{XiaomiCamera} crashes. In such a case, \textit{ChatApp} times out the request to \textit{XiaomiCamera}, leaving a reference to the requested picture set to null. When later the picture is used, a null pointer exception is thrown and \textit{ChatApp} stops working.}
\smallskip

In such a scenario, the in-vivo testing component will:
\begin{enumerate}
	\item recognize the configuration as \emph{unknown} (in fact it does not appear in the feature model depicted in Figure~\ref{fig:chatapp_fm});
	\item execute the available in-vivo tests, possibly retrieved from a testing server, to check if \textit{ChatApp} works properly with the camera app \textit{XiaomiCamera};
	\item expose a failure of \textit{ChatApp} (null pointer exception);
	\item report the failure, and the configuration that triggered it, to the developers.	
\end{enumerate}


\section{The \tool Framework} \label{sec:framework}

The \tool framework provides the architecture and a reference implementation that developers can exploit to add in-vivo testing capabilities to their mobile applications. In this context, we assume that the application under test (AUT) has been designed to support in-vivo testing. Note that this assumption does not necessarily imply that the AUT has been conceived for in-vivo testing from scratch, but rather that the AUT has been at some point extended with the minimal set of features required to support the in-vivo testing process. 


\subsection{Functional and Non-Functional Requirements} \label{sec:requirements}

The functional requirements reported below distinguish the functionalities that must be implemented by the in-vivo framework directly, and the functionalities that must be provided by the AUT by implementing interfaces defined in the framework.

\emph{FR-TestSpace}: The framework should be able to read a configuration model and a set of tested configurations from a persistent storage. 

\emph{FR-ActualConf}: The framework should expose an interface (\texttt{getConfiguration}) to be implemented by the AUT, by which it can determine the \emph{configuration} of the execution environment in which the AUT is deployed and is operated, as well as an interface by which the AUT can inform the framework of a new/updated configuration (\texttt{sendConfiguration}, \texttt{updateConfiguration}). 

\emph{FR-CheckConf}: The framework should identify the \emph{configuration} of the execution environment as \emph{tested},  \emph{untested}, or \emph{unknown}.

\emph{FR-TestExec}: The framework should be able to retrieve and execute in-vivo/ex-vivo tests for the \emph{untested} configurations. 

\emph{FR-TestGen}: The framework should be able to generate in-vivo/ex-vivo tests for \emph{unknown} configurations, possibly without, but if necessary with, manual intervention. 

\emph{FR-SelfHeal}: The framework should expose an interface to be implemented by the AUT, by which it can activate 
failure prevention/self-healing mechanisms in the presence of failing in-vivo/ex-vivo test executions. 

\emph{FR-Isolation}: The framework must ensure isolation of the in-vivo test executions, so that they do not interfere with regular operation of the AUT and do not have side effects (e.g., modify the persistent data of the user).

\smallskip
In addition to the functional requirements, we also identified a small set of relevant non-functional requirements that an in-vivo framework should satisfy.

\emph{NFR-PerfChecking}: The framework should not impose unacceptable levels of performance overhead when monitoring and checking the test configurations on the deployed AUT.

\emph{NFR-PerfTesting}: The framework should not impose unacceptable levels of performance overhead when running tests on the deployed AUT.

\emph{NFR-Network}: The network data usage (overhead) due to the communication between client and server components of the framework should be acceptable.

\emph{NFR-Energy}: The energy consumption due to the execution of the framework should be low.

\emph{NFR-Privacy}: The framework must ensure privacy of the client user when sharing information with the server.

\emph{NFR-Security}: The framework must ensure security of the client user in handling resources. 

Note that, although in this work we focus on mobile applications, our set of requirements are general and can be applied to many different contexts, including desktop and client-side Web applications.

\subsection{\tool Architecture} \label{sec:architecture}

\begin{figure}[tbh]
\begin{center}
 \includegraphics[scale=0.55, trim=1cm 8.5cm 10cm 0.5cm]{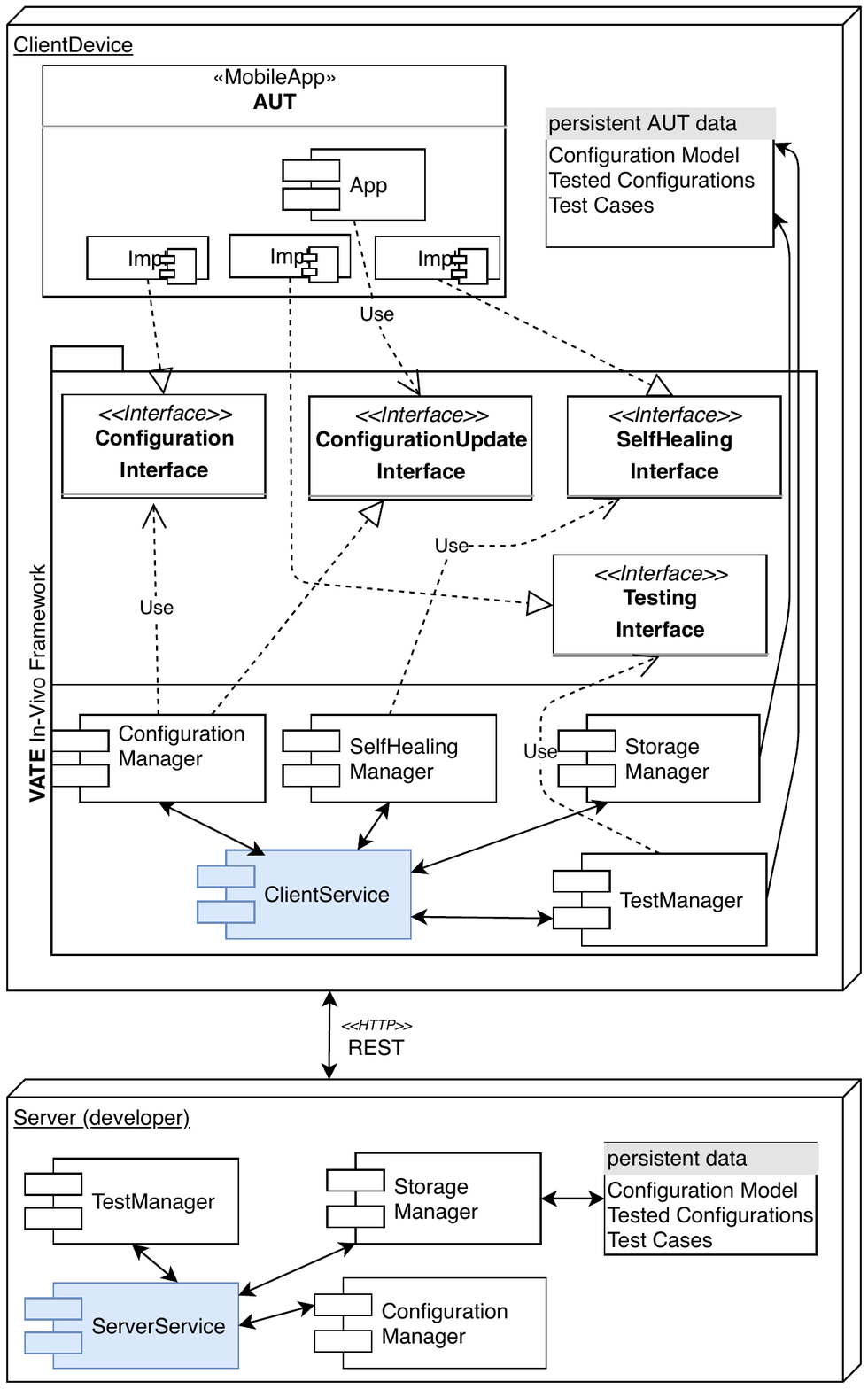}
\end{center} 
 \caption{\tool Architecture}
 \label{fig:framework}
\end{figure}

In order to satisfy the identified requirements, we designed the client-server architecture shown in Figure~\ref{fig:framework}. The client runs in the devices of the users and manages the in-vivo testing process local to the application under test (AUT). In fact, the client-side includes both the mobile app under test (AUT) and the \tool in-vivo framework, which is further organized in two layers, a layer of interfaces implemented by the AUT and a layer of managers responsible for both the in-vivo testing process and the interactions with the server. The server runs remotely and controls the in-vivo testing process by interacting with all the devices augmented with the in-vivo framework. When necessary, for instance when test cases cannot be conveniently executed in-vivo, the server-side can also run ex-vivo test cases. 

\vspace{-0.5cm}
\subsection{Client-Side Components} \label{sec:client}
There are four client-side components orchestrated by the ClientService, which offers the same entry point for all  components.

The \emph{Configuration Manager} is responsible for monitoring configurations, which consist of the hardware and software settings that may influence the behavior of the AUTs. This is done partially autonomously and partially in collaboration with the AUTs. In fact, the Configuration Manager can autonomously extract information about the hardware and the configuration of the operating system available in the client device. However, the Configuration Manager cannot access application-specific data without the collaboration and authorization of the AUT. For instance, the Configuration Manager may neither know where the app preference files are located nor have the right to access these files. To overcome this issue, the AUT must implement the \emph{Configuration Interface}, which is a read-only interface used by the Configuration Manager to extract a representation of the current configuration of the app. 

The Configuration Manager may simply query the interface when needed. However, a more efficient process may also allow the AUT to notify the Configuration Manager that the current configuration of the app has been modified. This is supported by the \emph{ConfigurationUpdate Interface}, which is defined in the framework and implemented in the AUT so as to generate notifications.

The Configuration Manager has also the responsibility to trigger test case execution and notify the existence of unexpected configurations to the server. In fact, every time a configuration is extracted, it is compared to both the Configuration Model, which is a representation of the possibly huge space of all the possible configurations, and the Tested Configurations, which is a representation of the configurations globally validated so far. 
The comparison of the current configuration to the configuration model and to the tested configurations can produce three possible results, as follows.
\begin{definition}[Tested Configuration]
A \textit{tested} configuration is a configuration that is valid according to the Configuration Model and is included in the Tested Configurations, which means it has been exercised either in pre-release testing or in-vivo testing.
\end{definition}

\begin{definition}[Untested Configuration]
An \textit{untested} configuration is a configuration that is valid according to the Configuration Model and is not included in the Tested Configurations, which means it has never been exercised, either in pre-release testing or in-vivo testing.
\end{definition}

\begin{definition}[Unknown Configuration]
An \textit{unknown} configuration is a configuration that is not valid according to the Configuration Model.
\end{definition}

When a tested configuration is discovered, it means that the current configuration has been already validated and nothing is done by the Configuration Manager. When an untested configuration is discovered, the Configuration Manager asks the Test Manager to validate it by running the in-vivo test cases. When an unknown configuration is discovered, the Configuration Manager notifies the server of the incompleteness identified in the Configuration Model, expecting to receive in the future a new version of the Configuration Model where the incompleteness has been fixed.

\smallskip


The \emph{Test Manager} is the component responsible for running the in-vivo testing process, reporting the results to the server, and updating the test suite available locally.

When the Test Manager is triggered to validate an untested configuration, it first checks with the server if the configuration is also untested globally. If the current configuration was already tested by another client, the server responds with an updated representation of the tested configurations and the process stops. Otherwise, if the configuration is globally untested, the Test Manager activates the available isolation mechanisms and runs the in-vivo test cases. The results of the testing process and the tested configuration are reported to the server, which can update the set of tested configurations.

\smallskip

The \emph{Storage Manager} is a simple component responsible for storing and updating the persistent data that characterize the in-vivo testing process: the configuration model, the tested configurations, and the in-vivo test cases. The Configuration Manager and Test Manager interact with the Storage Manager when these entities have to be retrieved or updated.

\smallskip
The \emph{Self-Healing Manager} is responsible for activating countermeasures when failures are detected. Some of these countermeasures might be activated treating the AUT as a black-box. However, the most sophisticated strategies may require collaboration from the AUT. To support the latter case, the architecture exposes a \emph{SelfHealing Interface}, to be implemented by the AUT to facilitate self-healing.

\section{The Android \tool Framework} \label{sec:implementation}

The architecture described in Section~\ref{sec:framework} is general and can be instantiated in multiple contexts using different technologies. One of the most relevant use cases for the \tool framework is for sure the Android ecosystem, where apps must work correctly in very heterogenous environments characterized by different hardware resources, different operating systems, and different user preferences whose combinations are impossible to test exhaustively~\cite{Wei:Fragmentation:TSE:toAppear}. We thus implemented a version of the \tool framework specifically for the Android ecosystem. 

In our first definition of the Android framework, we focus on the core functional and non-functional capabilities, leaving the implementation of the self-healing capabilities and the implementation of advanced mechanisms for  security and privacy for the future. Our implementation of the framework is publicly available~\cite{tool}. We describe the capabilities concerning configuration management, testing and isolation below.



\subsection{Configuration Management} \label{sec:configuration}

We specify the set of the possible configurations using a feature model~\cite{foda1990}. The feature model is obtained semi-automatically. Part of the configuration space is the same for every app, such as the part of the model that represents the hardware and operating system where an app can be executed. This part can be conveniently specified manually almost once for all. However there are also a number of app-specific settings that may affect the behavior of the apps. The volume of these settings is often significant. For instance, the feature model representing the configuration of \emph{Amaze File Manager}, one of the apps used in our experiment, contains 185 features (131 primitive and 54 compound)  resulting in a large configuration space whose size is in the order of $10^{12}$. Another app, \emph{RedReader}, has 567 features (461 primitive and 106 compound) and 4 constraints, resulting in a  configuration space of size in the order of $10^{58}$.   
Manually producing a feature model that represents so many elements might be prohibitive and error prone. To address this challenge, we designed a technique that can automatically extract the feature model corresponding to the preferences that appear in target preference files. Our approach essentially maps the Android Preference hierarchy into a feature model. To this end, we devised an appropriate feature model relation for each type of preference available in Android. In particular, \texttt{PreferenceCategory} and \texttt{PreferenceScreen} become abstract features, and the individual Preference items under them become their children in the feature model. For example, the Android preference \texttt{ListPreference} (a setting in which the user can select one of the available values) becomes a compound feature where each of the values is a primitive feature, all joined as \texttt{Alternative} features. Similarly, we have identified the appropriate mapping for the most common preference types in Android: \texttt{CheckBoxPreference, ListPreference, MultiSelectListPreference, PreferenceCategory, SwitchPreference}. We also introduce suitable heuristics for preference types where a direct mapping to feature model relations is not appropriate. In particular, for \texttt{EditTextPreference} we consider only two options: 'default value' and 'custom value'. Similarly, for numeric preference types, we introduce three options: 'negative', 'zero', 'positive'. Clearly such heuristics could be easily improved or replaced by customized categories depending on the nature of the app under investigation. For generic app preferences where we could not determine their types from their declaration, \tool may  not be able to map them automatically into the feature model. Such preferences require manual investigation to determine their type. 

The two feature models, the one representing the app-independent configurations and the one representing the app-dependent configurations, are merged into a single feature model that is used to \changed{guide} the in-vivo testing process. To manipulate the feature models, we make use of the FAMILIAR framework~\cite{DBLP:journals/scp/AcherCLF13}. In particular, we use FAMILIAR's simple syntax when generating feature models corresponding to Android preferences. Once generated, we transform the feature models in the FAMILIAR format into the SPLOT\footnote{\url{http://www.splot-research.org/}} format for easy manipulation and visual inspection/editing via the FeatureIDE\footnote{\url{http://www.featureide.com/}} plugin for Eclipse. Hence, while generating the entire feature model manually is quite difficult, the models automatically generated by our mapping could  be easily inspected and edited by the engineer as appropriate. 

While it would be possible to handle the representation of the tested configurations similarly, using FAMILIAR, we noticed that repeated application of the merging functionality of FAMILIAR produces complex models, involving a large number of constraints, eventually slowing down the check for tested configurations. Hence we developed an alternative, more compact, representation of the tested configurations using a tree-based representation. 

In order to retrieve the current configuration of a client-device, our Android implementation exploits two different mechanisms: the app-independent configuration is retrieved by the framework autonomously, without requiring any interaction with the AUT, while the framework interacts with the Configuration Interface to retrieve the content of the  preference files that were used to generate the app-dependent part of the feature model.

\subsection{In-Vivo Testing}
The In-Vivo testing process is performed entirely on the client-side and consists of a set of test cases that are executed to test the untested configurations discovered in the field. Test cases can be unit, integration and system test cases. 

We implemented the unit and integration test cases with JUnit. Since the existing system testing technologies are not designed to run from the device, we modified Espresso~\cite{Espresso} so that Espresso test cases can be launched and the results collected entirely from the device. 

Since test cases must be able to stimulate the AUT, the testing interface implemented by the AUT must include a method to launch the test cases, otherwise the in-vivo testing process may violate the security policies of the device. The implementation of this method is almost always the same and does not need to be designed ad-hoc for every target app. 

\subsection{Isolation}
Test case execution should be performed with minimal intrusiveness with respect to user activity. To achieve memory isolation, we exploit Managed Profiles~\cite{managedProfiles}, which are designed to support corporate environments (with corporate apps) on private employee devices. A managed profile represents an ideal technical solution for in-vivo testing, because it supports isolation and sand-boxing. 

{\em Isolation}: When an app is installed in a managed profile, it shares no data with the same app installed in the regular user profile (they are assigned distinct linux user-ids). Thus, the act of testing an app on the testing profile  does not affect the end-user data in the regular user profile.

{\em Sand-boxing}: The profile manager can dynamically install and remove an app from/to a managed profile (e.g., before/after running the test suite) and it can also dynamically grant and revoke permissions to apps in the managed profile, thus limiting the side effect of testing, such as information leakage.

The \tool framework defines an in-vivo testing profile where the AUT is copied the first time the in-vivo testing process is triggered. In-vivo testing happens within the in-vivo testing profile, thus it produces no side effects on the actual app used by the users and the user data. If, in addition to testing the app under the same hardware and operating system configuration, the app must be tested with the same software configuration and user preferences, the two copies of the app can be designed to communicate through an intent and exchange configuration data. Operations to achieve isolation with respect to external services, if any, must be implemented in the Testing Interface.      

Unit and integration testing can be performed transparently for the user. However, system testing requires taking  control of the screen and would thus be intrusive with respect to user activities. One way to mitigate such issue is to activate in-vivo testing when the screen is about to be locked: the Test Manager runs a single in-vivo test before allowing the screen to freeze. If more tests are to be executed for a given configuration, they will be run one at a time whenever new screen lock requests occur. Moreover, if the same configuration is observed in multiple devices, in-vivo tests are distributed (by the Server-Side Component of \tool) among different devices, hence reducing the impact on each single user. 

\section{Empirical evaluation} \label{sec:case-studies}

\subsection{Research questions}
The main goal of our experiments is to evaluate the performance impact of \tool. We also evaluated the level of automation achieved by our tool in the reverse engineering of feature models. Hence, we formulate the following research questions:

\begin{itemize}
\item \textbf{RQ\textsubscript{1}:}
\textit{What parts of the feature models were reverse engineered automatically by \tool and what parts required manual intervention?}
\item \textbf{RQ\textsubscript{2}:}
\textit{What is the overhead introduced by \tool when monitoring normal executions?}
\item \textbf{RQ\textsubscript{3}:}
\textit{For how long does \tool preempt the usage of the end-user device during in-vivo test execution?}
\end{itemize}


\subsection{Subjects}
\label{subsec:subjects}
We picked case study apps from those already used by PreFest~\cite{Lu:Pan:Zhai:Zhang:Li:2019} to test configuration-related programming defects. So, these are apps whose behaviour is known to change upon preference modification. Moreover, these apps are open source, so their source code is available for us to integrate \tool, and we can reuse test cases (generated by PreFest) that are meant to reveal programming defects related to user preferences. 

Since the integration with \tool requires some manual effort, we could not consider the full list of apps. 
Apps have been filtered, first of all, to discard those running on a very old version of Android (i.e., minSDK$<$14), which are incompatible with our framework. The remaining apps have been sampled, by selecting only one app per  domain, to maximize diversity in our data set.

The final \changed{set of apps is} listed in Table~\ref{table:apps}. For each app (name in first column), the table reports its size in Mb (second column) and the number of classes (third column). Moreover, for each app, the table reports the size of the configuration space (fourth column) and the number of test cases generated by PreFest (fifth column). 

\begin{table}[htb]
\centering
\caption{\changed{Set of} apps.}
\label{table:apps}
\begin{tabular}{l|rrrr}
\hline
App & Size (Mb) & Classes & Config. space & PreFest tests \\
\hline
Amaze & 11.9 & 219 & $10^{12}$ & 210 \\
Forecastie & 3 & 24 & $10^{10}$ & 300 \\
Materialistic & 4.05 & 136 & $10^8$ & 210 \\
Redreader & 5.61 & 255 & $10^{58}$ & 240 \\
Timber & 9.97 & 164 & $10^6$ & 240 \\
Uhabits & 4.52 & 209 & $10^6$ & 300 \\
\hline
\end{tabular}
\end{table}

\subsection{Procedure and metrics}
To answer  \textbf{RQ1} we investigate the proportion of app preferences automatically mapped to the configuration feature model by \tool. For all the subject apps, we applied \tool to build the feature model from the app preferences. For each app, we counted the number of preferences that \tool was able to directly map to the feature model (\emph{Direct}), the number of features mapped by applying manually defined, yet general heuristics available in \tool (\emph{Heuristics}), and those \tool was not able to map automatically (\emph{Unsupported}).


To answer \textbf{RQ2}, we conducted an experiment on Firebase test-lab\footnote{https://firebase.google.com/products/test-lab/}, a paid service provided by Google to test apps in the cloud. This service provides detailed analytics and resource consumption for the app under test.

To quantify the overhead introduced by \tool, we run \emph{two versions of each app}, (i) the original app as it comes just after compilation; and (ii) the app manually integrated with \tool, which monitors the configuration of the app and sends it to the server. We scheduled the monitoring service to make sure that a check is performed during the execution scenario subject to measurement.

We used Firebase test-lab to measure the following \emph{performance metrics}:
\begin{itemize}
\item {\em CPU Load:} Firebase test-lab monitors the CPU consumption and measures it constantly during execution. We compute the CPU load as the average CPU consumption during the entire execution of the test;
\item {\em Memory:} The maximum amount of memory used by the device, measured in MB;
\item {\em Network:} The total amount of data exchanged via the network interface, measured in KB.
\end{itemize}

{\bf Execution scenarios.} We want to measure the overhead due to our framework as it is perceived by an end-user in typical execution scenarios. Thus, we started from the app description, available in the official app store, and we identified the app main functional requirements described there (usually listed as bullet points at the beginning of the textual description). We defined an execution scenario for the first four functional requirements. We manually executed each scenario on the app, and we recorded it as an Espresso test case, using the corresponding feature in Android Studio.

Eventually, we obtained four Espresso test cases for each app, that represent four execution scenarios related to four main functional requirements. They allow us to collect metrics on typical user interactions.

The measurement experiment has been conducted on distinct \emph{devices} available in Firebase test-lab. Devices are shown in Table~\ref{table:rq2-devices}: device name in the first column and API version in the second column. The table also reports their number of cores (third column), the CPU frequency (fourth column) and the available memory (fifth column). They range from low-end (Moto E5 Play) to high-end (OnePlus 6T) devices.

\begin{table}[htb]
\centering
\caption{Devices used to measure \tool performance overhead.}
\label{table:rq2-devices}
\begin{tabular}{l|rrrr}
\hline
\textbf{Device} & \textbf{API} & \textbf{Cores} & \textbf{CPU Freq} & \textbf{RAM} \\
\hline
Moto E5 Play & 26 & 4 & 1.4 GHz & 2 GB\\ 
Google Pixel 2 & 28 & 8 & 4x2.35 GHz + 4x1.9GHz & 4 GB\\ 
OnePlus 6T & 28 & 8 & 4x2.8 GHz + 4x1.7GHz & 6 GB\\ 
\hline
\end{tabular}
\end{table}

The \emph{measurement procedure} consists of deploying each version of each app in Firebase test-lab, together with the execution scenarios in the form of Espresso test cases. Then, test cases are executed and metrics  are collected. To minimize random error, each test case has been executed 10 times.

The number of repetitions of each test case is based on the contrasting goals of (i) minimizing the experiment cost, in fact, Firebase test-lab \changed{charges} a cost proportional to the  execution time in  each device; and (ii) maximizing the accuracy of the measurement, by repeating it many times. The number of repetitions was calibrated on a first app and then used in all the apps, as follows.

Test cases of the first app (i.e., {\em Amaze}) have been initially \changed{executed} a large number of times in the two configurations. Then, for each test, we compared the collected metric values between the two configurations. 
We used these data to compute the power\footnote{We used the function {\em pwr.anova.test} from the {\em pwr} package available in R.} that the Anova test would achieve when used to reveal statistical significance in the performance difference with and without \tool. We, eventually, estimated how many repetitions would have been required for our statistical test to have a significant power (i.e. probability of committing type-II error $<20\%$ or power $>0.8$). The value of 10 repetitions resulted to be the best choice. 


In order to answer \textbf{RQ3}, we run the test cases of the case study apps on a physical device
and we measure their execution time. Test case execution time corresponds to the time needed by VATE to perform in-vivo testing, when the device is about to be locked. 


We relied on the test suites provided by Lu et al.~\cite{Lu:Pan:Zhai:Zhang:Li:2019}, which consist of the benchmark used to evaluate the PreFest tool, available from the PreFest  repository\footnote{https://github.com/Prefest2018/Prefest}. 

It is worth noting that PreFest tests are written in Python and run with Appium\footnote{http://appium.io/}, whereas \tool supports Espresso test cases written in Java. So, we developed a small program transformation module that translates PreFest tests to Java, to run them as Espresso tests.

To collect realistic time data, this experiment was conducted on an actual device connected to our computer, a Huawei Nexus 6P smartphone,  running Android 8.1.0.



\subsection{Experimental Results}

\begin{figure*}[tb]
	\includegraphics[width=0.98\textwidth]{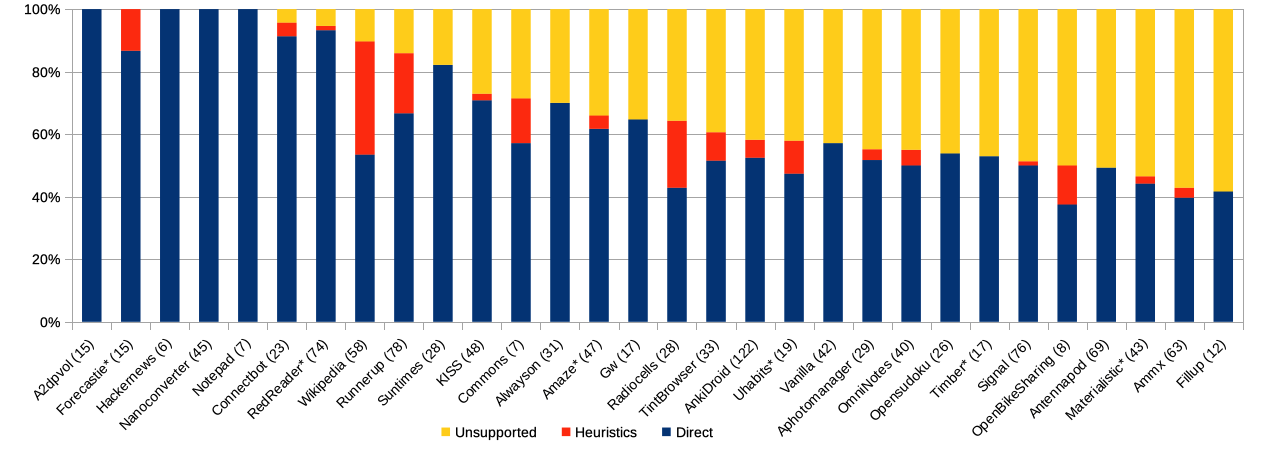}
	\caption{Proportion of preferences mapped by VATE to feature model (Direct, Heuristics) and those requiring manual intervention (Unsupported), shown in each bar as percentages. The total number of preferences in each app is shown in parentheses on the x-axis. The apps annotated with an asterisk (*) are used for the experiment of RQ2.}
	\label{fig:pref2fm_perc_barplot}
\end{figure*}

\textbf{Mapping of preferences to feature model.}
\changed{Given the high level of automation of the analysis, for this research question we considered a wider set of case studies than the one described in Section~\ref{subsec:subjects}: 30 apps instead of just 6.}
Figure~\ref{fig:pref2fm_perc_barplot} shows the proportions of preferences reverse engineered by \tool directly, via heuristics, or unsupported. The total number of preferences in each app is shown in parentheses next to the app name of the x-axis. The percentage of preferences automatically mapped by \tool to features of the target feature model ranges from 42\% (Fillup) to 100\%, $\approx$70\% on average. We can also notice that the large majority of  preferences translated by \tool are of \emph{Direct} type.

For those preferences whose type \tool is not able to determine automatically, manual inspection is required. For \textit{generic} preferences (\texttt{Preference}) and \textit{custom} preferences \textit{defined by the developer}, \tool is not able to determine automatically what options such preferences represent, and hence it is unable to map them to the feature model. Furthermore, some preferences are used to simply display information and do not actually represent any setting of the app (e.g., to display 'About' information). With manual inspection, the types of those preferences could be identified by looking at the declaration of the preferences in the xml files of each app and/or by running the app (e.g., in an emulator) and  observing what the preference actually represents. We did such an inspection for one of the apps (\emph{Amaze}) and it took us a couple of minutes to determine whether a generic preference declared in the xml file actually represents an app setting or not, and if it did, what type of setting it was. For the developer of the app this process should be even faster -- almost immediate. Furthermore, the construction of the configuration feature model is a task performed only once per app, with minor updates upon software evolution, if new settings are introduced or old ones are deleted.

%
%
%

\begin{center}
	\fbox{
		\begin{minipage}{0.95\linewidth}
		\begin{small}
			\textbf{In response to RQ1:} \tool was able to automatically reverse engineer, on average, about 70\% of the preferences. For those preferences whose type was not determined automatically by \tool, manual inspection required a short time (on average a couple of minutes).
		\end{small}
		\end{minipage}
	}
\end{center}
{\bf Analysis of \tool overhead}. The results of the experiment with Firebase test-lab are shown in Table~\ref{table:rq2-results}. For each app (first column) in each device\footnote{The app {\em Uhabits} is missing because was not compatible with the {\em moto-e5-play} device, so the corresponding line is missing.} (second column), the table reports the results for the metrics CPU, Memory, and Network. The {\em clean} columns report the average values collected on the original app, while the {\em \tool} columns report the values for the version with configuration monitoring. These columns report the average metric value for 10 executions of 4 test cases (thus 40 executions per column). The $\Delta\%$ columns report the metric increase (in percentage) due to \tool. The {\em \#sig} columns report the number of test cases for which the difference between the two versions (with and without \tool) are statistically significant according to the Wicoxon test. 

\begin{table*}[htb]
\centering
\caption{Results of the measurement in Firebase test-lab. Clean and \tool columns contains the amount of resources (CPU, memory and network) usage, the $\Delta\%$ column contains the percentage of overhead introduced by \tool and the \#sig column is the number of scenarios whose P-value is less than 0.05 (out of 4 scenarios)}
\label{table:rq2-results}
\begin{tabular}{ll|rrrr|rrrr|rrrrr}
\hline
&  & 
\multicolumn{4}{c|}{\textbf{CPU ($\%$)}} & 
\multicolumn{4}{c|}{\textbf{Memory (MB)}} & 
\multicolumn{5}{c}{\textbf{Network (KB)}} \\
\textbf{App} & \textbf{Device} & 
\textbf{Clean} & \textbf{\tool} & \textbf{$\Delta\%$} & \textbf{\#sig} &
\textbf{Clean} & \textbf{\tool} & \textbf{$\Delta\%$} & \textbf{\#sig} &
\textbf{Clean} & \textbf{\tool} & \textbf{$\Delta$} & \textbf{$\Delta\%$} & \textbf{\#sig} \\
\hline
Amaze & moto-e5 & 6.35 & 6.70 & 5.4 & 4 & 49.15 & 49.84 & 1.4 & 4 & 3.17 & 9.15 & 6.0 & 188.4 & 3 \\ 
   & pixel2 & 1.80 & 1.87 & 3.9 & 4 & 105.81 & 106.13 & 0.3 & 1 & 3.61 & 8.87 & 5.3 & 146.0 & 3 \\ 
   & oneplus-6t & 1.46 & 1.50 & 3.2 & 3 & 113.53 & 114.18 & 0.6 & 0 & 2.05 & 4.85 & 2.8 & 136.5 & 4 \\ 
\hline
  Forecastie & moto-e5 & 4.86 & 4.95 & 1.8 & 2 & 29.66 & 29.77 & 0.4 & 3 & 92.11 & 96.77 & 4.7 & 5.1 & 2 \\ 
   & pixel2 & 1.22 & 1.26 & 2.8 & 2 & 63.50 & 63.71 & 0.3 & 0 & 95.52 & 99.21 & 3.7 & 3.9 & 1 \\ 
   & oneplus-6t & 1.07 & 1.09 & 1.4 & 0 & 72.57 & 72.78 & 0.3 & 1 & 92.93 & 93.25 & 0.3 & 0.3 & 1 \\ 
\hline
  Materialistic & moto-e5 & 11.49 & 10.79 & -6.1 & 0 & 50.65 & 48.93 & -3.4 & 0 & 2499.49 & 2065.53 & -434.0 & -17.4 & 0 \\ 
   & pixel2 & 3.12 & 3.10 & -0.7 & 0 & 107.63 & 102.39 & -4.9 & 1 & 2992.51 & 2873.78 & -118.7 & -4.0 & 0 \\ 
   & oneplus-6t & 2.85 & 2.77 & -2.7 & 0 & 122.12 & 113.30 & -7.2 & 0 & 3213.63 & 2919.72 & -293.9 & -9.1 & 0 \\ 
\hline
  RedReader & moto-e5 & 14.75 & 15.00 & 1.7 & 0 & 37.54 & 37.93 & 1.0 & 2 & 24154.90 & 23130.27 & -1024.6 & -4.2 & 0 \\ 
   & pixel2 & 4.06 & 4.14 & 1.8 & 0 & 75.33 & 76.40 & 1.4 & 2 & 32940.51 & 32645.19 & -295.3 & -0.9 & 0 \\ 
   & oneplus-6t & 3.75 & 3.79 & 1.0 & 0 & 83.68 & 84.78 & 1.3 & 3 & 33384.89 & 30707.90 & -2677.0 & -8.0 & 0 \\ 
\hline
  Timber & moto-e5 & 4.90 & 4.99 & 2.0 & 2 & 44.49 & 43.55 & -2.1 & 0 & 23.61 & 26.93 & 3.3 & 14.1 & 2 \\ 
   & pixel2 & 1.47 & 1.48 & 0.8 & 0 & 87.52 & 88.54 & 1.2 & 3 & 30.12 & 31.45 & 1.3 & 4.4 & 1 \\ 
   & oneplus-6t & 1.24 & 1.27 & 2.6 & 1 & 104.43 & 104.96 & 0.5 & 1 & 18.56 & 20.35 & 1.8 & 9.7 & 3 \\ 
\hline
  Uhabits & pixel2 & 1.07 & 1.08 & 1.3 & 1 & 70.27 & 71.17 & 1.3 & 4 & 0.31 & 4.35 & 4.0 & 1295.0 & 4 \\ 
   & oneplus-6t & 0.89 & 0.89 & -0.0 & 0 & 76.51 & 76.60 & 0.1 & 1 & 0.00 & 2.66 & 2.7 & -- & 4 \\ 
\hline
\end{tabular}
\end{table*}

As can be seen from Table~\ref{table:rq2-results}, the CPU load due to \tool is quite limited,  always below 6\% and in the majority of the cases no significant difference can be observed between the load with and without our framework. The largest number of significant cases is observed for the app {\em Amaze}, because this app is a file manager and its test cases involve deterministic scenarios, such as file creation and compression. Conversely, {\em Materialistic} and {\em Redreader} (a news and a Reddit reader \changed{app, respectively}) involve nondeterministic scenarios, because different executions of the same test case could load different news/articles. The presence of noise  dominates any effect due to the overhead of \tool, which becomes no more observable. Correspondingly, there is no statistical significance and some deltas are negative due to random fluctuations caused by noise.
 
The memory overhead is also quite limited, always below 1.5\%, and in many cases the difference due to \tool is not significant (i.e., {\em \#sig}$<4$). Negative overheads of {\em Materialistic} and {\em Timber} are due to random errors in measuring memory consumption, again because of nondeterministic execution scenarios, and they are not statistically significant. Test cases for {\em Materialistic} run on {\em pixel2} exhibit a negative average memory overhead that is significant in one test. This delta is negative because it is the average over four test cases. However, we manually checked that the overhead for the single significant test case was actually positive.

The network cost of \tool accounts for the app configuration sent to the sever, so its {\em absolute} value is constant and predictable. However, the {\em relative} overhead reported in  column $\Delta\%$ is a percentage of increase. Thus, when the original app (i.e., {\em clean} column) has a limited network usage, the relative increase appears to be large. This is the case of apps such as {\em Amaze} and {\em Uhabit} which involve almost exclusively offline scenarios. On the other hand, on scenarios that involve network usage, the \tool relative overhead is negligible. This is the case, for example, of app {\em Forecastie}, which fetches remote weather forecast data. Thus, only for the network metric, we also report the {\em absolute} increase in an additional $\Delta$ column. 
The negative overhead observed in {\em Materialistic} and {\em Redreader} is not significant and is caused by random noise. In fact, network traffic depends on the size of news and articles randomly fetched by these apps.
When significant, the network increase is always limited to few kilobytes, in the worst case 6KB for {\em Amaze} running on {\em moto-e5}.



\begin{center}
	\fbox{
		\begin{minipage}{0.95\linewidth}
		\begin{small}
			\textbf{In response to RQ2:} The overhead introduced by VATE while monitoring app configurations is quite low: the CPU load increase is negligible in most of the cases and always $<6\%$, the memory overhead is $<1.5\%$, and the network overhead is below 6KB \changed{per client-server interaction}.
		\end{small}
		\end{minipage}
	}
\end{center}
\textbf{Analysis of the in-vivo test execution time.} 
With this study, we measure the time necessary to run a system test case with \tool. These results are important to understand the impact of \tool on the users in various scenarios. For instance, if \tool is configured to opportunistically run a system test case before a device is locked, these measures provide an estimate of the locking delay that might be experienced by users who allow the execution of system tests in their device.  

As reported in Table \ref{tab:espresso_device}, the mean execution time for the Espresso tests is lower than 5 seconds, with a standard deviation (s.d.) which is between 2 and 3 seconds. The statistical error of the estimated mean is lower than 5\% in all cases except Materialistic, which has the highest relative s.d. (i.e., s.d./mean ratio). 

The capability to run a system test in few seconds confirms the possibility to apply \tool in real Android devices. For instance tests could be feasibly executed in a 5 seconds session whenever the screen lock is triggered. The next test can be launched at a future time, when another screen lock occurs, but just in case it has not already been executed meanwhile on another device.

\begin{table}[htb]
	\centering
	\caption{Execution time for the Espresso tests running on a Nexus 6P device}
	\label{tab:espresso_device}
	\begin{tabular}{lrrr}
		\toprule
		\bf App & \bf Mean (s) & \bf SD & \bf Error\\
		\midrule
		Amaze File Manager & 4.04 & 2.29 & 0.01 \\ 
		Forecastie & 3.24 & 2.39 & 0.04 \\
		Materialistic & 2.52 & 2.72 & 0.24 \\
		Redreader & 4.28 & 2.59 & 0.04 \\
		Timber & 3.11 & 2.18 & 0.05 \\
		\bottomrule
	\end{tabular}
\end{table}

\begin{center}
	\fbox{
		\begin{minipage}{0.95\linewidth}
		\begin{small}
			\textbf{In response to RQ3:} The time spent to run a single test is consistently under 5 seconds. We consider this an acceptable time, e.g., in a scenario where each test runs independently in a 5 seconds session whenever the screen lock is triggered. 
		\end{small}
		\end{minipage}
	}
\end{center}

\textbf{Reproducibility:} We make \tool openly available~\cite{tool}, together with a replication package including features models, test cases, and pointers to the subject apps.

\subsection{Threats to validity}

\textbf{External validity:} Our results are based on 6 apps and on the test cases defined for them by the authors of PreFest~\cite{LuPZ0L19}. Such test cases have not been designed explicitly for in-vivo execution, although they are designed to exercise preference-dependent portions of the application code. While our results may not generalise to different apps and to test cases explicitly designed for in-vivo execution, we have chosen the  benchmark available from the closest related work, dealing with preference-based testing.

\textbf{Internal validity:} The measures of overhead were obtained while running test scenarios that are supposed to mimic the typical usage scenarios of the  apps. While we did our best to produce such scenarios based on the main functionalities advertised for each app under test, we cannot exclude that the overhead may change under different usage conditions.

\section{Related work} \label{sec:rel-work}

While to the best of our knowledge \tool is the first framework that supports in-vivo testing for mobile (specifically, Android) applications, there are previous works that deal with related problems. In particular, the problems of in-vivo monitoring and isolation have been already considered, though not in the mobile domain. Preference-based testing for mobile applications is also related to our work, in particular to coverage of the feature combinations described in our configuration models.  

\textbf{Techniques to isolate in-vivo test execution}
Several techniques~\cite{GonzalezSanchezPG09,HuningJSSE2010,LahamiKJ15,MurphyKVC09} have been proposed to support isolation during in-vivo testing. \textit{Duplication} (also called \textit{Cloning})~\cite{GonzalezSanchezPG09,HuningJSSE2010,MurphyKVC09} consists of cloning the execution state (e.g., by forking a parallel process~\cite{MurphyKVC09}) and executing in-vivo tests on the cloned execution state, hence ensuring that there is no interference with the end user execution of the application (in-memory side effects are prevented, but of course other side effects on persistent storage are not dealt with). Another proposed isolation mechanism is \textit{Test mode execution}~\cite{BartoliniJSS2011,GreilerPESOS2010,KawanoOM89,LahamiSCP2016,ZhuTSC2012}. It requires a way to differentiate between the execution of a component in normal operation mode vs. the testing mode. In the latter case, counter measures are taken to ensure that test mode execution does not affect the normal execution state (e.g., by tagging invocations and data with a test tag~\cite{KawanoOM89,LahamiKJ15}). Another clean and elegant solution consists of using a transactional memory~\cite{BobbaPACT2009}. Field tests can perform their operations within a transaction and at the end of their execution, such a transaction is rolled back and normal execution restarts exactly from the memory state where it was interrupted for in-vivo test execution. Other authors propose that developers write \textit{built-in tests}~\cite{SammodiSAC2011},  specifically designed for in-vivo test execution. It is then the developers' responsibility to ensure that such tests are side effect free. Differently from existing works, our solution to the isolation problem takes advantage of the managed profiles available in the Android platform (see Section~\ref{sec:framework}).

\textbf{Preference based testing}
Lu et al.~\cite{LuPZ0L19} showed that different preference configurations lead to different code coverage by the same test cases and that proper selection of which preference configurations to test can increase statement (resp. branch) coverage on average by 6.8\% (resp. 12.3\%). They also provide evidence that some (five, in their experiment) bugs require specific preference settings to be discovered. Such empirical results represent a major motivation for our work: when the configuration space grows and depends on environment/device/user specific settings, offline, pre-release testing is not enough to exercise the code and expose the faults that depend on such configurations. PreFest~\cite{LuPZ0L19} performs static code analysis to determine the code that is potentially data-dependent on user preferences and selects the test cases that can exercise such a code, along with the associated preference values. \tool resorts to in-vivo test execution to cope with the exponential growth of the possible configurations, as well as their unavailability during in-house testing.

\textbf{Combinatorial testing} Testing all valid configurations exhaustively before deploying an app is not feasible because the number of combinations grows exponentially with the number of features and because some combinations might require very specific hardware/software components. Combinatorial testing (e.g., pairwise testing)~\cite{CohenGMC03} offers a way to systematically explore such a large configuration space. However, by sampling a small representative fraction of all possible cases, it leaves several combinations untested. Some of them might be handled incorrectly at runtime, resulting in field  failures.

\textbf{Empirical studies on field failures}
Gazzola et al.~\cite{GazzolaMPP17} investigated the nature of field failures by analyzing the bug reports for five applications. They introduce the notion of ``field-intrinsic fault'', that is, a field fault that is inherently hard to detect in-house, before releasing the software. 
They also identify the  reasons why faults are not detected at testing time. 
They conclude that there is evidence of a relevant amount of faults that cannot be effectively addressed in-house and should be addressed directly in the field. Such findings represent an important motivation for the work presented in this paper.

\section{Conclusion and future work} \label{sec:concl}

We have presented \tool, a framework for in-vivo testing of Android apps, and we have measured its overhead on the end-user executions, showing that such overhead is compatible with in-field adoption of our approach. In our future work we plan to optimize configuration monitoring, making it adaptive and distributed. We also intend to investigate test case generation in response to newly discovered configurations. We are also considering other application domains for \tool, such as that of web applications.

\tool together with a replication package including features models, test cases, and pointers to the subject apps are available online~\cite{tool}.

\section*{Acknowledgements}
\footnotesize{
This work has been partially supported by the Italian Ministry of Education, University, and Research (MIUR) with the PRIN project GAUSS (grant n. 2015KWREMX); by the H2020 Learn project, funded under the ERC Consolidator Grant 2014 program (ERC Grant Agreement n. 646867); by the H2020 Precrime project, funded under the ERC Advanced Grant 2017 program (ERC Grant Agreement n. 787703).

We would like to thank Filip Ivanov Karchev for contributing to the implementation of the initial in-vivo prototype, in particular for engineering a solution for on-device execution of Espresso test cases and for contributing to the initial sketch of the in-vivo server.
}

\balance

\nocite{MurphyKVC09,GazzolaMPP17}
\bibliographystyle{abbrv}
\bibliography{refs}

\end{document}